\documentclass[aps,twocolumn,showpacs]{revtex4}
\usepackage{dcolumn}
\usepackage{graphicx}
\usepackage{amsmath}
\usepackage{amsfonts}
\usepackage{amssymb}
\usepackage{psfrag}
\usepackage{wrapfig}
\usepackage{subfigure}
\usepackage{makeidx}
\usepackage{bm}
\usepackage{epsf}

\begin{document}

\title{Pair-Tunneling Induced Localized Waves in a Vector Nonlinear
Schr\"{o}dinger Equation}
\author{Li-Chen Zhao$^{1}$, Liming Ling$^{2}$, Zhan-Ying Yang$^{1}$, Jie Liu$^{3,4}$}\email{liu_jie@iapcm.ac.cn}
\address{$^1$Department of Physics, Northwest University, 710069, Xi'an, China}
\address{$^2$Department of Mathematics, South China University of Technology, Guangzhou 510640, China }
\address{$^3$Science and Technology Computation Physics Laboratory,
 Institute of Applied Physics and Computational Mathematics, Beijing 100088,
China}
\address{$^4$Center for Applied Physics and Technology, Peking University, 100084, Beijing,
China}
\date{August 14, 2013}
\begin{abstract}

We investigate the localized waves of the coupled two-mode nonlinear
Schr\"{o}dinger equations with a pair-tunneling term representing
strongly interacting particles can tunnel between the modes as a
fragmented pair. Facilitated by Darboux
 transformation, we have derived exact solution of nonlinear vector waves such as
 bright solitons, Kuznetsov-Ma
soliton, Akhmediev breathers and rogue waves and demonstrated their
interesting  temporal-spatial structures.
 The phase diagram that demarcates the parameter ranges  of  the nonlinear waves is obtained.
 Our  results have implications in such diverse fields as Bose-Einstein condensate, nonlinear fibers and super fluids.

\end{abstract}
 \maketitle

\emph{Introduction}---Vector nonlinear waves  such as vector bright
soliton (BS), vector Akhmediev breather (AB) and vector rogue waves
(RW), have recently become a topic of intense research in theory
because of their widespread applications
\cite{Pitaevskii,Baronio,Zhao,Lakshmanan,Bludov,Becker}. These kinds
of waves  appear in oceans,  atmosphere, optics, plasmas, as well as
in the quanta world of super fluids and Bose-Einstein condensates
(BEC) \cite{N.Akhmediev,Osborne}. Most studies  are based on the
coupled nonlinear Schr\"{o}dinger equations(NLS), in which coupling
between different modes (or components) are described by cross-phase
modulation(XPM) term. One mode  can then influence others through
imposing a phase  that is dependent of its instantaneous density
distribution. Thus, for the XPM-type coupling, the population or
particle number in each component is conserved. However, in
practical physical systems, the mode population  in each component
is not necessarily conserved. For instance, in microscopic particle
transport or light propagation, the particle (or light) in one well
(or mode) can transfer to another well (mode) through quantum
tunneling \cite{Li,Williams,Wadati,Qin,Lahini}. More interestingly,
for the BEC systems where the ultra-cold atoms behave coherently,
pair-tunneling (PT)  that the strongly correlated atoms can tunnel
back and forth as a fragmented pair in a double-well, were observed
in recent experiments \cite{pair,Meyer}. Behind dynamics can be
described by a two-mode  vector NLS that contain both XPM and PT
coupling terms. In such systems, particle population in each mode is
not conserved and nonlinear waves are expected to be more marvelous.

In this letter, we show that the two-mode vector NLS with both XPM
and PT couplings can be integrable and present exact solutions of
vector BS, vector Kuznetsov-Ma soliton (K-MS), vector AB and vector
RW using Darboux transformation method. These nonlinear waves
demonstrate marvelous temporal-spatial  structures. We
observe Josephson-like oscillation for vector K-MS solution.  For the
vector RW, the typical eye-shape structure of total density
distribution is found to  decompose into a hump density distribution of one
component and a two-valley density distribution of the other
component. We obtain the phase diagram that demarcates the parameter
ranges  of  the distinctive nonlinear waves and further demonstrate more
exotic  second-order nonlinear wave.

\emph{Integrable Model}--- With including tunneling effects,
dynamics of two-component systems can be  described by following
vector nonlinear Schr\"{o}dinger Equation in general\cite{Williams},

\begin{eqnarray}
i\left(
\begin{array}{c}
\Phi_{1t} \\ \Phi_{2t}\end{array}
 \right)
&=& H \left(
\begin{array}{c}
\Phi_{1} \\ \Phi_{2}\end{array}
 \right),
\end{eqnarray}
\begin{eqnarray}
H&=&\left(\begin{array}{cc}H_1^0+H_1^{MF}-\delta/2& \Omega
\\ \Omega^*& H_2^0+H_2^{MF}+\delta/2\end{array} \right),\nonumber
\end{eqnarray}
 where $\delta$ denotes the energy difference between the two components. The star means complex conjugation.
 $H_i^0=-\partial_x^2$ are free evolution.  $H_i^{MF}=g_{i,i} |\Phi_i|^2
 +g_{3-i,i}|\Phi_{3-i}|^2$ $(i=1,2)$ are the interactions accounting for self-phase
 modulation and XPM (incoherent coupling effect), represented by the first and second term  respectively in the righthand.
 $\Omega$ represents the tunneling term, denoting coherent coupling between the components.
 In most studies,  $\Omega$ is set to be zero  because it was usually believed that the presence of tunneling makes the systems become non-integrable\cite{Baronio,Zhao,Lakshmanan,Bludov}.
 While in a more recent work \cite{Li}, it was found that, when the tunneling term takes a specific form of delta function of $\Omega=\delta(x)$,
 a static vector BS solution can be derived with dropping the XPM term.  In the present work, we
 consider $\Omega=\beta \Phi_1^*
 \Phi_2$, denoting PT effects observed in recent BEC experiments \cite{pair,Meyer}, and
 the coherent energy coupling effects in a nonlinear birefringent
fiber \cite{G.P.Agrawal}. Moreover, the nonlinear XPM term is
present.

 With setting $\delta=0$, $g_{1,1}=g_{2,2}=\beta=-2$ and $g_{2,1}=g_{1,2}=-4$, the equation (1) can be rewritten explicitly as the following integrable coupled NLS,
 \begin{eqnarray}
i \Phi_{1t}+\Phi_{1xx}+2 |\Phi_1|^2 \Phi_1+4|\Phi_2|^2 \Phi_1+2
\Phi_2^2 \Phi_1^*&=&0, \\
i \Phi_{2t}+\Phi_{2xx}+2 |\Phi_2|^2 \Phi_2+4|\Phi_1|^2 \Phi_2+2
\Phi_1^2 \Phi_2^*&=&0.
\end{eqnarray}
It should be pointed that the integrable coupled NLS is distinctive from the well-known Manakov system \cite{Forest, Wright}, since the pair-tunneling effects are considered in the coupled model. In following, we will present the exact solutions of nonlinear waves such as vector BS, vector K-MS, vector AB, vector RW, and even
high-order ones,  and  demarcate the parameter boundary of these nonlinear waves.

\emph{Pair-tunneling induced localized waves}---With solving the
corresponding Lax-pair from the seed solution $\Phi_{10}=s
\exp{[i2s^2
 t]}$ $(s>=0)$ and $\Phi_{20}=0$,
the generalized vector nonlinear wave solution can be deduced  as
follows using Dauboux transformation method\cite{Matveev},
\begin{eqnarray}
\Phi_1&=& s\exp{[i 2 s^2t]} +\frac{2a}{|P_1|^2+|P_2|^2} P_1 P_2^*, \\
\Phi_2&=& \frac{2a}{|P_1|^2+|P_2|^2} P_1 P_2^*,
\end{eqnarray}
where \begin{eqnarray}
 P_1&=&  \frac{\sqrt{\tau+a
}}{\tau}\Psi_1-\frac{\sqrt{-\tau+a
}}{\tau}\Psi_2,\nonumber\\
P_2&=& -\left(\frac{\sqrt{-\tau+a }}{\tau}\Psi_1-\frac{\sqrt{\tau+a
}}{\tau}\Psi_2\right)
 \exp{[-i 2 s^2t]},\nonumber
\end{eqnarray}
and
 \begin{eqnarray}
 \Psi_1&=&  \exp{\left[\tau x
+i\left( \tau^2+2 a  \tau- a ^2+ 2 s^2 \right) t\right]} ,\nonumber\\
\Psi_2&=& \exp{\left[-\tau x+i\left( \tau^2-2 a \tau- a ^2+ 2 s^2
\right) t\right]},\nonumber
\end{eqnarray}
 with $\tau=\sqrt{a^2-s^2}$ ($a$ is an arbitrary real number).
 $a$ determines the initial nonlinear localized wave's
shape, and $s$ is related with the amplitude of each background for
localized waves.  In the following, we discuss properties of the nonlinear wave solution
 according to the two parameters.
\begin{figure}[htb]
\centering {\includegraphics[height=82mm,width=85mm]{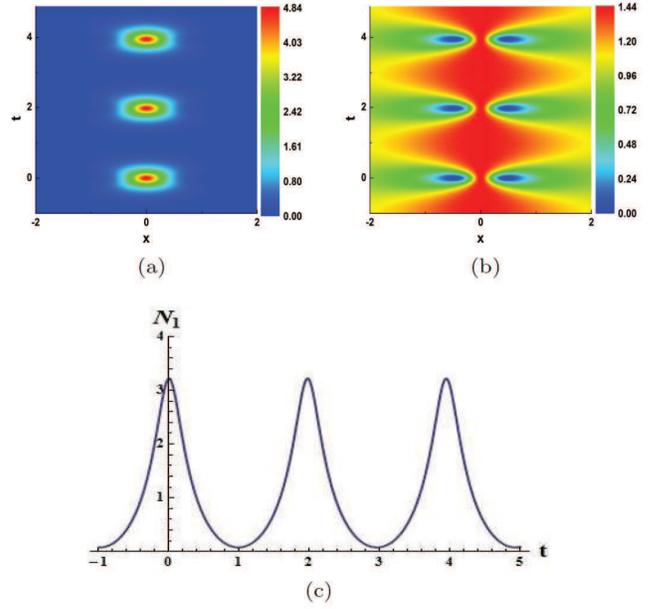}}
\caption{(color online)The evolution of vector K-MS which is
periodic in time: (a) the density evolution of component $\Phi_1$,
and (b) the density evolution of component $\Phi_2$. (c) The
evolution of all particles number in $\Phi_1$. It is shown that the
oscillation can be seen as nonlinear Josephson oscillation. The coefficients are $s = 1$ and $a=1.2$. }
\end{figure}

Case 1: When $s>0$ and $|a|>s$, the temporal spatial structures of
the solution is presented in Fig. 1. The densities  of both
components show breathing structure. The sum over them, i.e., total
density, is found to be analogous to the K-MS of scalar NLS
\cite{Kuznetsov,Ma}. As an extension, we term it as vector K-MS.
Interestingly, we observe periodic conversion between two
components. By calculating the whole particle number
 $N_1[t]=\int_{-\infty}^{\infty}
|\Phi_1|^2 dx$, we find that, the oscillation is not cosine or sine
type as in standard Josephson oscillation\cite{J}, shown in Fig. 1
(c). Instead, it has been modified by the nonlinear interactions,
similar to the nonlinear Josephson effects observed in BEC
experiments\cite{E}. The oscillation period  is obtained  as
follows,
\begin{equation}
T=\frac{\pi}{2 a\sqrt{a^2-s^2}}.
\end{equation}

\begin{figure}[!b]
\centering {\includegraphics[height=42mm,width=85mm]{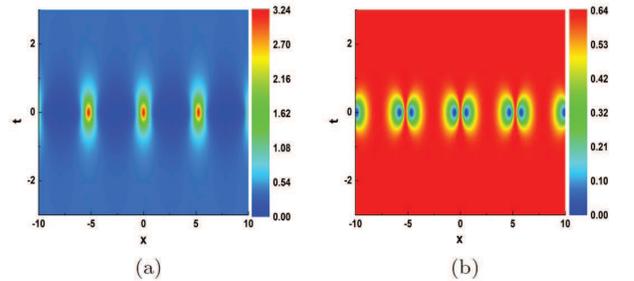}}
\caption{(color online) The evolution of vector AB: (a) the density
evolution of component $\Phi_1$, and (b) the density evolution of
component $\Phi_2$. The coefficients are $s = 1$, and $a=0.8$. }
\end{figure}

The amplitude of the oscillation can be calculated according to the
difference between the maximum and the minimum of particles number
in the first component. It reads
\begin{eqnarray}
\bigtriangleup N_1&=& \int |\Phi_1|_{max}^2-|\Phi_1|_{min}^2 dx
\nonumber\\
&=&\pi s ,
\end{eqnarray}
where $|\Phi_1|^2_{max}$ and $|\Phi_1|^2_{min}$ denote the maximum
value and minimum value of the density in time, respectively. It
indicates that the number of the particles involved in  the
tunneling is proportional to the amplitude of the nontrivial
background.

Case 2: When $s>0$ and $|a|<s$, the vector wave is periodic in
space and the behavior of  sum density distribution is analogous to the
AB of scalar NLS \cite{AB}.  Therefore, it is marked by vector AB as shown in Fig. 2.
It is seen that a hump train appears in the density distribution of the first component, while a
double-valley  train emerges for the second component. Each double-valley structure looks like a butterfly.
The temporal periodic oscillation is greatly suppressed so that
 the particle conversion between the two components only emerges in certain time range. The density period can be derived analytically,
\begin{equation}
S=\frac{\pi}{\sqrt{s^2-a^2}}.
\end{equation}

Case 3: When $|a|=s>0$, vector RW solution can be
derived from the general solution with taking a limit $\tau
\rightarrow 0$. The periodicity in both time and space is absent, as
 shown in Fig. 3. A double-valley structure appears in the density distribution of the second component $\Phi_2$, while, a hump structure appears in the density distribution of the first component $\Phi_1$. In the double-valley structure, the highest density emerges at the center between
the two valleys and its amplitude is equal to the background density.
We certify that the double-valley structure in component of $\Phi_2$ is quite
distinctive from usual RW structure for scalar NLS. However, the distribution of the
sum density of  both components is analogous  to the well-known Peregrine
solution of scalar NLS \cite{Peregrine}, which has been considered
as RW \cite{Kibler}. The above observation suggests that, in the presence of PT,  the typical eye-shape RW of total density is decomposed into a hump density distribution of $\Phi_1$
component and a double-valley density distribution of $\Phi_2$
component.

\begin{figure}[!t]
\centering {\includegraphics[height=42mm,width=85mm]{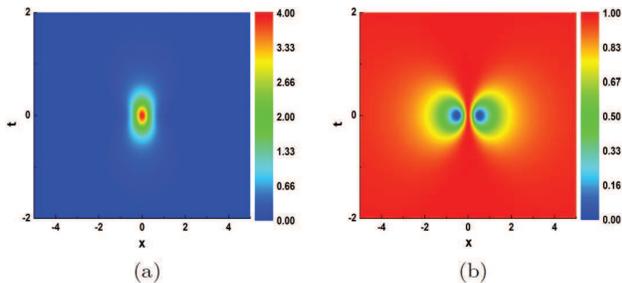}}
\caption{(color online) The evolution of vector RW: (a) the density
evolution of component $\Phi_1$, and (b) the density evolution of
component $\Phi_2$. The coefficients are $s =a=1$. }
\end{figure}
Case 4: When $s = 0$ and $a\neq 0$, the vector bright soliton(BS)
can be given directly from the general expression of the exact solution. In this case,
no tunneling emerges  between  two components. The vector solution is a trivial
combination of the well known scalar solitons of NLS, that is,  the
solitons' shape in the two components are identical.
\begin{figure}[!t]
\centering {\includegraphics[height=45mm,width=60mm]{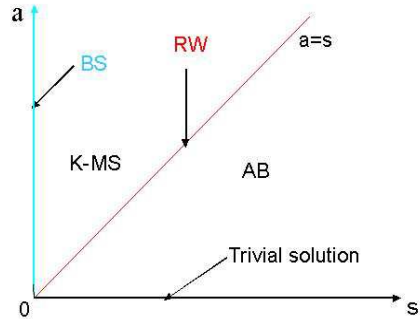}}
\caption{(color online) The phase diagram of nonlinear waves with
parameters $a$ and $s$. The BS denotes vector bright solitons, RW
denotes vector rogue waves,  K-MS denotes vector Kuznetsov-Ma
solitons which are periodic in time, and AB denotes vector Akhmediev
breathers which are periodic in space.}
\end{figure}

\begin{figure}[htb]
\centering {\includegraphics[height=85mm,width=85mm]{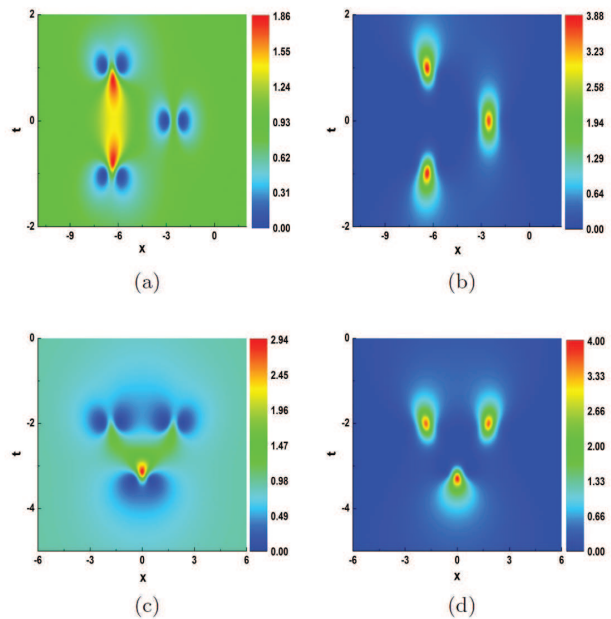}}
\caption{(color online) The evolution plot of the second-order RW
in the coupled system. (a) for one kind RW in $\Phi_1$ component,
(b) for the RW in $\Phi_2$ component correspondingly.  (c) for
another kind RW in $\Phi_1$ component, (d) for the RW in $\Phi_2$
component correspondingly. It is seen that the valleys in $\Phi_1$
correspond to the humps in $\Phi_2$ for pair-tunneling effects. }
\end{figure}

To summarize the above cases in parameter $a$ and $s$ space, we plot
an interesting phase diagram in Fig. 4. We just plot the first
quadrant with $s\geq 0,a\geq 0$, it can be extended to other
quadrants according to centric symmetry. On the $s$ axis, all
particles keeps staying in  $\Phi_1$ component, no population in
$\Phi_2$, corresponding to a trivial plane solution.

In addition to the above marvelous localized waves, high-order
nonlinear  solutions can also be obtained through Darboux
transformation method. As an example, we show the second-order RW
with the PT effects. The actual expression of the solution is quite
complicated and will be presented elsewhere. In Fig. 5, we plot the
second-order RWs for different cases, which shows more interesting
patterns emerged. We emphasize that the PT effects are essential in
generating the above exotic structures.

\emph{Application into two-component condensate
system}---Experimentally, vector soliton has been realized in
multi-component BEC systems \cite{Das,Engels}, and in a nonlinear
birefringent fiber \cite{Tang}. On the other hand, scalar AB, RW and
K-MS have also been observed very recently in single-mode nonlinear
fibers \cite{Kibler,Hammani2,Kibler2}. The PT induced localized
waves could be observed in the birefringent fiber or two-component
BEC systems through combining these experimental techniques.

\begin{figure}[!t]
\centering {\includegraphics[height=45mm,width=85mm]{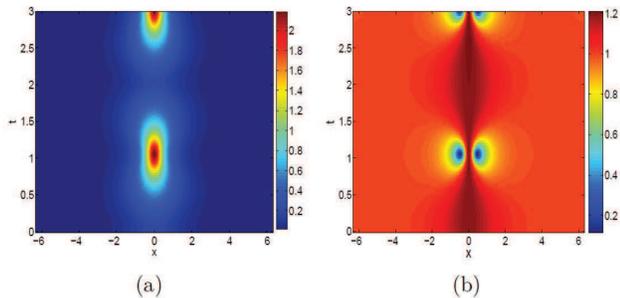}}
\caption{(color online) The numeric stimulation results for vector K-MS wave: (a) the density
evolution of component $\Phi_1$, and (b) the density evolution of
component $\Phi_2$. The initial conditions are given by Eq. (4) and (5) with $s=1$, $a=1.2$,
and $t=-1$. It is seen that the numeric evolution of localized waves is similar to the exact ones in Fig. 1. }
\end{figure}

 Let us
 consider a cigar-shaped condensate with two hyperfine
states, $\Phi_1$ and $\Phi_2$. For simplicity, we assume the initial
condensation occurring in the trapped state $\Phi_2$. State $\Phi_1$
is coupled to $\Phi_2$ by an RF or microwave field tuned near the
$\Phi_2 \rightarrow \Phi_1$ transition. The PT effects can be
realized by the RF field in the strong interaction regimes
\cite{pair,Meyer,Ballagh}. The total number of $\,^{87}Rb$ atoms in
the condensate is $N= 5\times 10^4$. $a_{i,j}(i,j=1,2)$ are s-wave
scattering lengths which can be adjusted by Feshbach resonance
technique. Setting $a_{1,2}=a_{2,1}=1.6\ nm$ and
$a_{2,2}=a_{1,1}=0.8 \ nm$, under mean-field approximation, the
s-wave scattering effective interaction strengths between atoms in
the same hyperfine state are $U_{j,j}=4 \pi \hbar^2 a_{j,j} /m$ ($m$
is the atom mass), and the scattering effective interaction
strengths between atoms in different hyperfine state are
$U_{j,3-j}=4 \pi \hbar^2 a_{j,3-j} /m$.
 When the interaction between atoms is
attractive and the PT coefficient is $N\cdot U_{1,1}$, the units in
axial direction and time are scaled to be $2.0\ \mu m$ and $1.0 \
ms$ respectively, the dynamics of the condensate with PT effects can
be described well by the Eq. (2) and (3). The exact solution can be
used to design proper initial density and phase conditions in the
two components for each localized wave presented here. For example,
from the K-MS initial condition, we would observe the nonlinear
Josephson oscillation between the two hyperfine states. From the
above results, we can know that the oscillation period of the K-MS
waves in Fig. 1 will be $1.97\ ms $. Since the density and phase can
be manipulated well in BEC systems \cite{Pitaevskii}, the localized
waves could be observed in the two-component condensate system.

In summary, the PT induced localized waves, such as BS, AB, K-MS,
and RW, are investigated exactly in the two-component system. We stimulate the localized waves by numeric calculation from the initial condition given by the exact solution. It is found that the
localized waves can be excited in the system. As an example, we show the stimulation results for vector K-MS localized wave in Fig. 6, which correspond to the exact ones in Fig. 1. Our
obtained solutions contribute to better control and understanding of
localized wave phenomena in a variety of complex dynamics, ranging
from optical communications, to Bose-Einstein condensates, and
four-wave mixing systems.

L.C. Zhao is grateful to Dr. Hui Cao for helpful discussions. This
work is supported by the National Fundamental Research Program of
China (Contact 2011CB921503, 2013CB834100), the National Science
Foundation of China (Contact Nos. 11274051, 91021021).

\end{document}